\documentclass[11pt,a4paper]{article}

\usepackage[margin=25mm]{geometry}
\usepackage[numbers,sort&compress]{natbib}
\usepackage[utf8]{inputenc}
\usepackage[T1]{fontenc}
\usepackage{hyperref}
\usepackage{url}
\usepackage{booktabs}
\usepackage{amsfonts}
\usepackage{amsmath}
\usepackage{nicefrac}
\usepackage[protrusion=true,expansion=false]{microtype}
\usepackage{xcolor}
\usepackage{graphicx}
\usepackage{float}
\usepackage{multirow}

\begin{document}

\title{GAGI: A Gini-Adjusted GDP-per-Capita Index for
Distribution-Aware Macroeconomic Welfare Monitoring}

\author{%
  Sivasathivel Kandasamy \\
  Independent Researcher \\
  \texttt{sivasathivel@yahoo.com}
}

\date{}
\maketitle

\begin{abstract}
Gross domestic product per capita is the default lens through which
governments and international bodies track the economic prosperity and consequences of events 
artificial-intelligence-driven automation, yet GDP is blind to two
first-order determinants of lived prosperity: how income is distributed
and how far it stretches after inflation. Inequality-adjusted income
measures are themselves not new --- the Atkinson index, the
inequality-adjusted Human Development Index, and the World Bank's
shared-prosperity premium all combine growth and distribution in some
form. What is missing from the macroeconomic monitoring toolkit specifically is not
a welfare \emph{concept} but an operational \emph{monitoring trigger}: a
statistic minimal enough to compute annually from public data, transparent
enough to audit without modelling assumptions, and normalised so that
year-on-year, cross-country \emph{change} --- the quantity a regulator
needs to act on --- is legible. We assemble such an instrument, the
\emph{Gini-Adjusted GDP per Capita Index} (GAGI): a reproducible,
publicly computable formulation that rescales each country's GDP per
capita by its inequality-adjustment factor $(1-G)$ and its price level,
normalised to a 2010 baseline. GAGI is a general-purpose welfare
index --- not inherently specific to AI automation --- applicable
wherever welfare-adjusted prosperity needs tracking; we apply it here
to examine the welfare effects of AI-driven automation in the G7 over
2010--2026 on a timescale that matches policy decision cycles. Applying
GAGI to the G7 economies over 2010--2026, we
show that welfare-adjusted prosperity has diverged persistently and
increasingly from headline GDP growth, that the divergence widens sharply
after 2022, temporally coincident with --- though not, on this evidence
alone, demonstrated to be caused by --- after effects of COVID and the acceleration of generative-AI
deployment,
and that the resulting trajectories separate into qualitatively distinct
regimes --- automation without absorption (United States) and structural
stagnation without automation (Italy, Japan) within the core G7 sample,
contrasted with automation \emph{with} absorption in Nordic comparator
economies drawn from outside the G7 (Denmark, Sweden) --- that are
clearly separable in GAGI--Gini space despite being indistinguishable in
GDP space. We document the rise of AI-attributed labour displacement in
the United States --- the only G7 economy for which comparable
employer-attributed layoff data currently exist, where employer-cited AI
layoffs rose $13\times$ between 2023 and 2025 --- alongside the
cross-country productivity--wage divergence (a 14-percentage-point cumulative gap between
labour productivity and real wage growth in high-income countries,
1999--2024), and middle-skill employment hollowing, and we present
three scenario-based projections of G7-average GAGI through 2035 that
bound the welfare stakes of unconstrained automation. We argue that GAGI
is a necessary complement to GDP-based monitoring: any macroeconomic
monitoring instrument that tracks only aggregate output will
systematically miss the distributional harm that automation can cause
even while reported growth remains strong. GAGI is designed as a
plug-in empirical proxy for the Economic Stability ($E$) component of
the Human Utility Factor (HUF), a companion welfare-optimisation
framework for identifying optimal parameters of alignment between
automation and macro-socio-economic wellbeing \citep{huf2026}.
\end{abstract}

\section{Introduction}
\label{sec:intro}

A government, a regulator, or a firm deciding how aggressively to deploy
automating AI systems needs an answer to a deceptively simple question:
\emph{is the resulting productivity translating into broadly shared human
welfare, or is it being captured by a narrowing share of the population
while the majority falls behind in real terms?} GDP per capita --- the
statistic most commonly invoked in this debate --- cannot answer the
question, because it is a population-level average that is blind to both
distribution and purchasing power. A country can report record GDP growth
while the median household's inflation-adjusted, inequality-adjusted
living standard stagnates or falls.

This is not a hypothetical concern. Between 2022 and 2026, AI-attributed
layoffs in the United States rose more than thirteen-fold
\citep{challenger_2026,challenger_march2026}; global labour productivity
in high-income economies has outpaced real wage growth by fourteen
percentage points cumulatively since 1999 \citep{ilo2024wages}; and the
labour share of national income has fallen across most OECD economies
\citep{karabarbounis2014global,karabarbounis2024perspectives}. Each of
these facts is individually well documented. What is missing is a single,
reproducible, cross-country statistic that aggregates them into a
trackable welfare signal --- one that a regulator could plausibly compute
from public data and use as an early-warning indicator, the way central
banks track inflation or unemployment.

To be clear about what is, and is not, novel here: Gini-adjusted income
measures are a well-established family, and we make no claim to have
invented a new welfare-economic construct (\S\ref{sec:discussion} situates
GAGI relative to the Atkinson index, the inequality-adjusted Human
Development Index, and the World Bank's shared-prosperity premium). Our
contribution is narrower and more operational: a specific
\emph{formulation and use-case} --- minimal data requirements, a
fixed-baseline normalisation, and full auditability --- engineered so
that the statistic can function as a general welfare-monitoring trigger
rather than a retrospective research artefact. GAGI is not inherently
specific to AI automation: it is applicable wherever welfare-adjusted
prosperity needs tracking across populations, regions, or time periods.
The present paper demonstrates this generality by applying it to the
particular question of AI-driven automation in the G7. We propose such an instrument:
the \textbf{Gini-Adjusted GDP per Capita Index} (GAGI). GAGI rescales GDP per capita by an inequality-adjustment
factor (one minus the Gini coefficient) and by the consumer price index,
then normalises the result to a fixed base year. The construction is
deliberately simple --- every input is published annually by the World
Bank, the IMF, or national statistical agencies --- so that the index can
be computed and audited by any interested party without proprietary data
or modelling assumptions. Section~\ref{sec:metric} defines GAGI formally.
Section~\ref{sec:results} applies it to the G7 economies, 2010--2026, and
documents five empirical patterns: a persistent and widening GAGI--GDP
wedge (\S\ref{sec:s_country}), the acceleration of AI-attributed labour
displacement in the United States --- the only G7 economy with comparable
employer-attribution data, so this evidence is reported as U.S.-specific
rather than generalised to the G7 (\S\ref{sec:s_layoffs}) --- the
cross-country productivity--wage divergence (\S\ref{sec:s_wages}), the
separation of national trajectories by policy regime, illustrated with
Nordic comparator economies drawn from outside the core G7 sample
(\S\ref{sec:s_gini}), and middle-skill employment hollowing
(\S\ref{sec:s_polarization}). Section~\ref{sec:projections} presents
scenario-based projections of GAGI through 2035 and argues that the
choice of governance regime --- not the pace of automation per se ---
determines which trajectory a country follows. Section~\ref{sec:discussion}
discusses GAGI's relationship to existing welfare metrics, its limitations,
and its intended role as a plug-in empirical proxy for the Economic
Stability ($E$) component of the Human Utility Factor (HUF), a
constrained-optimisation framework for identifying the parameters that
align automation with macro-socio-economic wellbeing, developed in a
companion paper \citep{huf2026}.

\section{The GAGI Metric}
\label{sec:metric}

For country $k$ and year $t$, we define
\begin{equation}
  \mathrm{GAGI}_{k,t}
  \;=\;
  \frac{\bigl(1 - G_{k,t}\bigr)\;\times\;\mathrm{GDP\text{-}pc}_{k,t}}
       {\pi_{k,t}}
  \;\bigg/\;
  \frac{\bigl(1 - G_{k,2010}\bigr)\;\times\;\mathrm{GDP\text{-}pc}_{k,2010}}
       {\pi_{k,2010}},
  \label{eq:gagi}
\end{equation}
where $G_{k,t}\in[0,1]$ is the Gini coefficient \citep{worldbank2025pip},
$\mathrm{GDP\text{-}pc}_{k,t}$ is GDP per capita in constant 2015 USD
\citep{worldbank2025wdi}, and $\pi_{k,t}$ is the consumer price index
(2010\,=\,100) \citep{worldbank2025wdi}. Dividing by the 2010 baseline
converts the statistic from a level into a measure of \emph{change},
which is what makes cross-country comparison meaningful: countries differ
enormously in absolute GDP per capita and in absolute Gini, but the
question of whether welfare-adjusted prosperity is rising or falling
\emph{relative to a fixed reference point} is comparable across very
different economies. A GAGI value above 1.0 indicates that
inequality-adjusted, inflation-corrected prosperity has improved relative
to 2010; a value below 1.0 indicates deterioration.

\paragraph{On dividing constant-price GDP by the CPI.} A careful reader
will note that $\mathrm{GDP\text{-}pc}_{k,t}$ is published in constant
2015 USD --- i.e., already deflated once, by the (output-side) GDP
deflator --- and may ask whether dividing by $\pi_{k,t}$, a
consumption-side price index, double-counts inflation. It does not,
because the two indices price \emph{different baskets}. The GDP deflator
reflects the prices of goods and services a country \emph{produces}
(domestic output, including investment goods, government consumption, and
exports, and excluding imports); the CPI reflects the prices of the
basket households \emph{actually buy} (consumption goods, including
imported consumer goods, housing, energy, and food). These baskets can
diverge sharply --- as they did during the 2021--2023 terms-of-trade and
energy-price shock, when import and consumer-energy prices rose far
faster than domestic-output prices across most G7 economies --- and the
gap between them is itself a cost-of-living effect that a production-side
real-GDP figure cannot register. Re-expressing constant-price GDP per
capita in CPI terms therefore corrects for the wedge between what an
economy \emph{produces} in real terms and what its households can
\emph{actually purchase} with the income that production generates ---
which is the welfare-relevant quantity GAGI is designed to track, not a
re-statement of the same inflation adjustment. That said, we record this
as a considered modelling choice rather than a settled convention. A
useful robustness check would recompute GAGI under a single-deflation
specification (nominal GDP per capita divided by the CPI alone, with no
separate constant-price step); we have not yet run that comparison and do
not present a result for it here, but flag it as a natural target for a
revised version of this paper. We would expect the GDP-deflator/CPI wedge
to be a second-order correction relative to the Gini-driven divergence
that dominates the patterns in Figure~\ref{fig:country_subplots}, but we
state that as an expectation, not a finding.

GAGI is intentionally minimal. It uses three inputs, all published
annually for essentially every country in the world, and it makes no
assumption about the causal mechanism connecting automation to outcomes.
This is a deliberate design choice: a metric that requires
country-specific modelling cannot serve as a cross-country monitoring
instrument, because its outputs would not be comparable. GAGI sacrifices
mechanistic detail for auditability and reproducibility --- properties
that we argue are prerequisites for any statistic intended to function as
a governance trigger rather than merely a research artefact. In a
companion paper \citep{huf2026} we show that GAGI is also a close
empirical analogue of the \emph{Economic Stability} sub-function $E(\cdot)$
in the Human Utility Factor (HUF), a derived welfare metric for
identifying optimal alignment between automation and
macro-socio-economic wellbeing; the present paper develops GAGI
independently as a standalone empirical instrument and documents what
it reveals about the 2010--2026 period.

\section{Empirical Patterns, G7 Economies, 2010--2026}
\label{sec:results}

This section presents evidence across five dimensions: per-country GAGI
trajectories (\S\ref{sec:s_country}), AI-attributed labour displacement
(\S\ref{sec:s_layoffs}), the productivity--wage divergence
(\S\ref{sec:s_wages}), inequality trends and the policy-regime effect
(\S\ref{sec:s_gini}), and middle-skill employment hollowing
(\S\ref{sec:s_polarization}).

\paragraph{Data vintage and provenance.} Because the sample period runs
into the still-unfolding present, we are explicit throughout about which
figures rest on finalised data and which do not, using four tiers: (i)
\emph{observed} --- finalised World Bank, OECD, and BLS series, available
through 2023--2024 depending on the series
\citep{worldbank2025pip,worldbank2025wdi}; (ii) \emph{preliminary} ---
2024--2025 figures from recent IMF/World Bank releases that remain open to
revision \citep{imf2025weo}; (iii) \emph{extrapolated} --- 2026 values
inferred from partial-year (Q1) data under a constant-growth assumption
\citep{imf2026weo,challenger_march2026}; and (iv) \emph{scenario/projected}
--- the model-based 2025--2035 trajectories of
Section~\ref{sec:projections}, which are explicit illustrations of policy
stakes and not forecasts. Figures mark tiers (ii)--(iii) with dotted
lines and hatched bars, and Section~\ref{sec:projections} carries its own
disclaimer. The reader should weight the tiers accordingly: the paper's
load-bearing empirical claims rest on tier (i); tiers (ii)--(iv) indicate
direction and order of magnitude and should be read as provisional,
extrapolated, or hypothetical respectively, not as established fact. We
also flag grey-literature sources (employer reports, consultancy
estimates) explicitly wherever they appear, since they carry different
evidentiary weight than peer-reviewed or official-statistics sources.

The central empirical finding is that the GAGI--GDP wedge is visible in
every G7 economy, that it has widened materially since 2022, and that the
trajectories separate into qualitatively distinct national patterns once
inequality is taken into account.

\subsection{Per-Country Welfare Trajectories}
\label{sec:s_country}

\begin{figure*}[t]
  \centering
  \includegraphics[width=\textwidth]{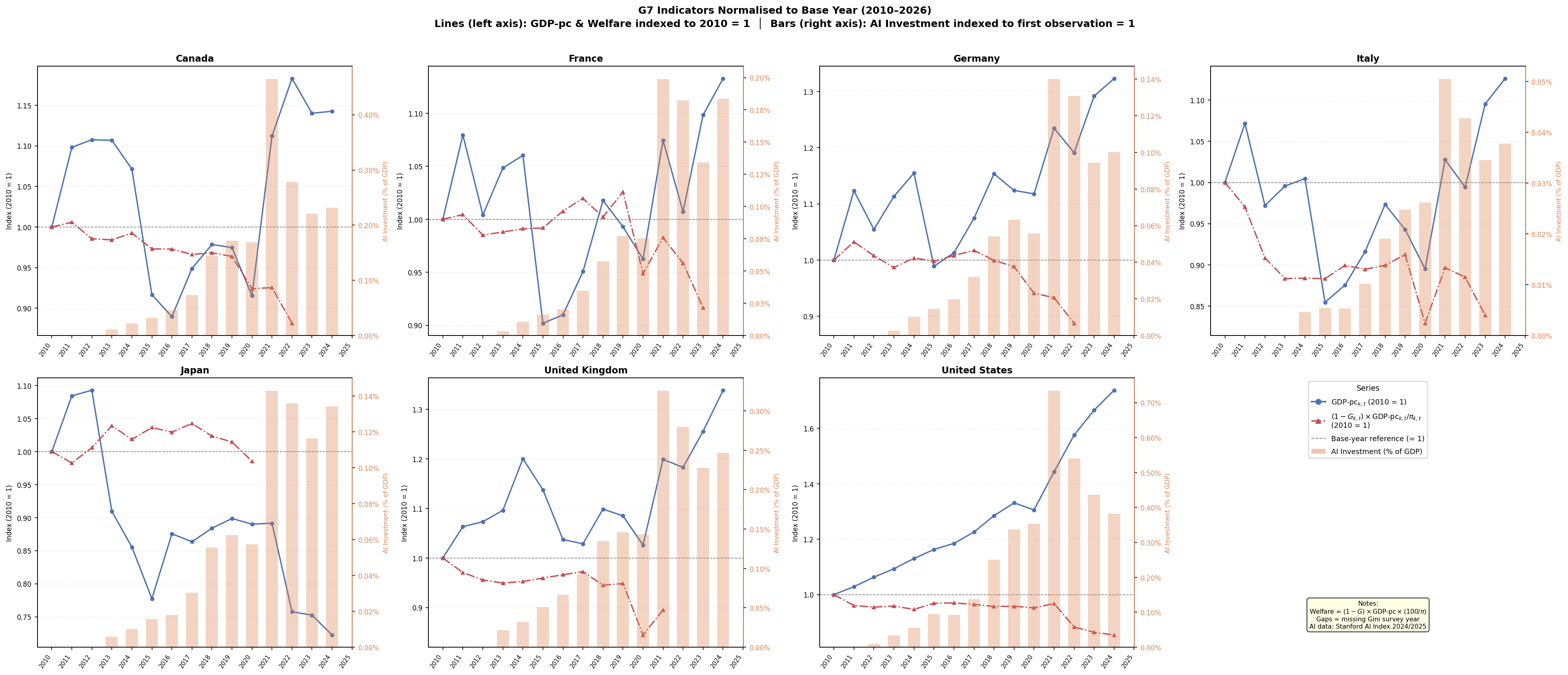}
  \caption{%
    \textbf{Gini-Adjusted GDP per Capita Index (GAGI), GDP per Capita
    Index, and Private AI Investment, G7 Economies, 2010--2026.}
    \textbf{Observation.} The GDP per capita index (blue line, left axis)
    consistently exceeds the GAGI index (red line, left axis) in every G7
    country throughout the period, with the gap widening post-2022,
    coincident with accelerating AI investment (bars, right axis,
    country-specific scale). GAGI\,=\,$(1{-}G)\times\mathrm{GDP}_{pc}/\pi$,
    normalised to 2010\,=\,1.0 (Eq.~\ref{eq:gagi}).
    \textbf{Interpretation.} The persistent divergence quantifies the
    distributional cost that raw GDP growth conceals: productivity gains
    are accruing faster than they reach households once rising inequality
    and inflation are accounted for. The United States shows the largest
    and most sustained gap; Japan and Italy exhibit near-flat GAGI despite
    low AI investment, exposing structural stagnation as a distinct
    welfare-failure mode that is unrelated to automation intensity.
    \textbf{Implication.} A monitoring regime that tracks only GDP --- as
    essentially all current macroeconomic monitoring instruments do
    --- will systematically underestimate the welfare cost of
    automation-driven distributional harm, and will do so silently, since
    headline growth figures continue to look healthy throughout.
    Sources: \citet{worldbank2025pip,worldbank2025wdi,%
    imf2025weo,imf2026weo,aiindex2024,aiindex2025}.%
  }
  \label{fig:country_subplots}
\end{figure*}

Figure~\ref{fig:country_subplots} presents GAGI (solid line), the GDP per
capita index (dashed line), and private AI investment as a share of GDP
(bars) for each G7 economy, 2010--2026. Values for 2025 are preliminary
IMF World Economic Outlook estimates \citep{imf2025weo}; values for 2026
are partial-year extrapolations from Q1 data \citep{imf2026weo}; both are
marked with dotted lines and hatched bars.

Three cross-country patterns emerge. First, in every G7 economy the GDP
per capita index lies above the GAGI index throughout the period: raw
economic growth consistently overstates welfare-adjusted prosperity once
inequality and inflation are accounted for. The wedge has widened in the
post-2022 period in all seven countries, coincident with the acceleration
of AI investment and the layoff patterns documented in
\S\ref{sec:s_layoffs}. Second, the United States shows the largest and
most sustained GAGI--GDP divergence, a pattern that co-occurs with --- though
the cross-sectional data here cannot establish that it is explained by ---
both its outsized AI-investment share and its comparatively weak
labour-market absorption institutions (\S\ref{sec:s_gini}). Third, Japan and Italy display near-flat GAGI
despite low AI investment, exposing structural stagnation as a distinct
failure mode: low automation does not by itself protect welfare when
underlying productivity growth is absent.

\subsection{AI-Attributed Labour Displacement: Evidence from the United
States, 2022--2026}
\label{sec:s_layoffs}

\emph{Scope note.} No economy in our G7 sample other than the United
States currently publishes a comparable employer-attributed displacement
series; the evidence in this subsection is therefore U.S.-specific and
should not be read as a G7-wide measurement of AI-driven displacement.
Where cross-country evidence exists, it takes the form of model-based
\emph{exposure} estimates (task shares that \emph{could} be automated),
which are a different and weaker kind of evidence than realised,
employer-attributed layoffs; we report both kinds below and are careful
to keep them distinct.

The Challenger, Gray \& Christmas Job Cut Reports provide the only
longitudinal time series of employer-attributed AI layoffs in the United
States \citep{challenger_2026}. AI was not tracked as a distinct cause
until May 2023; since then the series shows rapid escalation: 4,247
AI-cited cuts announced in 2023, rising to 12,742 in 2024 and 54,836 in
2025 --- a thirteen-fold increase over three years. In Q1 2026 alone,
27,645 additional cuts were attributed to AI, with AI overtaking all other
stated causes to become the leading reason given for U.S. layoffs in March
2026 \citep{challenger_march2026}. The cumulative total through March 2026
stands at 99,470 AI-cited announcements, roughly 3.5\% of all announced
cuts in that span.

\begin{table}[h]
\caption{AI-attributed layoffs and tech-sector displacement (United
  States, 2022--Q1 2026). \emph{AI-cited cuts}: employer-stated reasons,
  Challenger, Gray \& Christmas \citep{challenger_2026,challenger_march2026};
  category introduced May 2023. \emph{Tech layoffs}: headcount from
  Layoffs.fyi \citep{layoffsfyi2026}. All figures are \textbf{announced}
  separations, not realised; AI attribution reflects employer
  self-reporting. $\dagger$\,Q1 2026 only (through 31 March 2026).}
\label{tab:layoffs}
\centering
\footnotesize
\setlength{\tabcolsep}{4pt}
\begin{tabular}{@{}lrrrr@{}}
\toprule
\textbf{Year} & \textbf{AI-cited} & \textbf{\% of total}
  & \textbf{All cuts} & \textbf{Tech layoffs} \\
\midrule
2022 & ---     & ---    & 363,824   & 164,969 \\
2023 & 4,247   & 0.6\%  & 721,677   & 264,220 \\
2024 & 12,742  & 1.7\%  & 761,358   & 152,922 \\
2025 & 54,836  & 4.5\%  & 1,206,374 & 124,201 \\
2026$^\dagger$ & 27,645 & 12.7\% & 217,362 & 95,878 \\
\midrule
\textbf{Cumul.\ 2023--Mar\,26}
  & \textbf{99,470} & \textbf{3.5\%}
  & \textbf{2,906,771} & \textbf{637,220} \\
\bottomrule
\end{tabular}
\end{table}

Displacement is not uniform across the workforce. \citet{atkinson2026young}
find that workers aged 22--25 in the most AI-exposed occupations
experienced a 13\% employment decline between November 2022 and December
2025, driven by lower job-finding rates for new entrants rather than
layoffs of incumbents. Simultaneously, \citet{davis2026wages} document
that nominal weekly wages in high-AI-exposure industries rose 16.7\%
since fall 2022 versus 7.5\% economy-wide --- a bifurcation consistent
with early-stage skill-premium acceleration documented in prior
technology-displacement cycles \citep{autor2003skill}.

\begin{figure}[h]
  \centering
  \includegraphics[width=\linewidth]{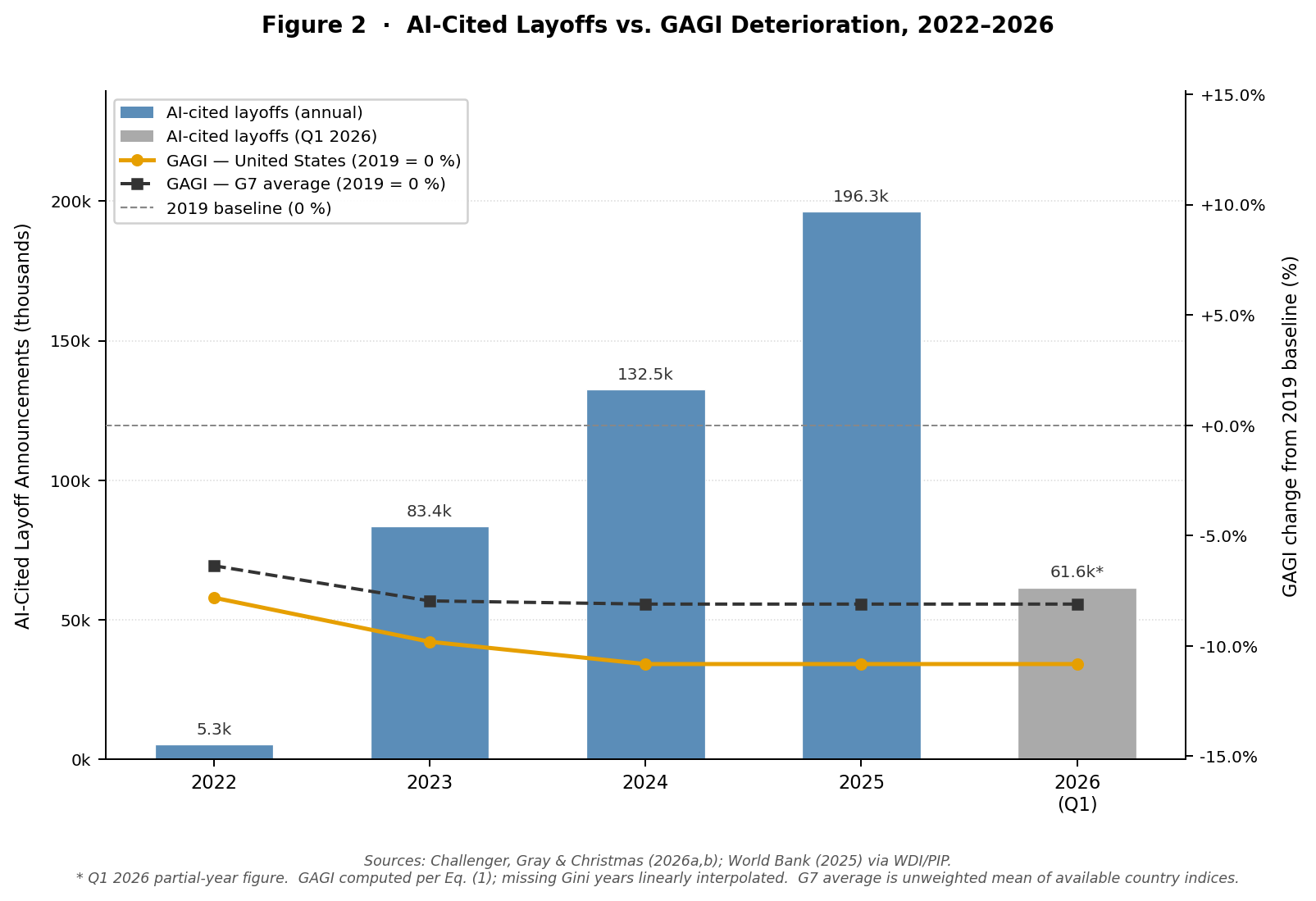}
  \caption{%
    \textbf{AI-Cited Layoffs vs.\ GAGI Deterioration, 2022--2026.}
    \textbf{Observation.} AI-attributed layoffs (bars, left axis;
    Challenger, Gray \& Christmas \citep{challenger_2026,challenger_march2026})
    grew from 4,247 in 2023 to 54,836 in 2025, while the U.S. GAGI index
    (amber, right axis) declined relative to the G7 average (grey dashed)
    over the same period.
    \textbf{Interpretation.} The co-movement of rising AI-attributed
    displacement and declining welfare-adjusted prosperity is consistent
    with --- though this time series alone cannot establish --- a channel
    in which the productivity gains from automation accrue disproportionately
    to firms while displacement costs fall on workers and households.
    Whichever causal channel is ultimately responsible, the co-movement
    documented here is precisely the kind of wedge between aggregate growth
    and distributed welfare that GAGI is designed to register.
    \textbf{Implication.} A regulator monitoring only GDP would observe
    record revenues --- consistent with a growing economy --- while the
    welfare-relevant consequence, the GAGI decline, accumulates
    undetected on the other axis.
    Sources: \citet{challenger_2026,challenger_march2026,%
    worldbank2025pip,worldbank2025wdi}.%
  }
  \label{fig:layoffs_gagi}
\end{figure}

Figure~\ref{fig:layoffs_gagi} links this displacement trajectory directly
to the GAGI deterioration measured in Figure~\ref{fig:country_subplots}.

Cross-sectoral exposure estimates corroborate the scale of the underlying
shift. \citet{eloundou2024gpts} estimate that 80\% of U.S. workers have at
least 10\% of their tasks exposed to large language models.
\citet{briggs2023ai} report task-automatable shares of 46\% for office and
administrative support, 44\% for legal occupations, and 37\% for
architecture and engineering. \citet{cazzaniga2024genai} (IMF Staff
Discussion Note 2024/001) find that 60\% of jobs in advanced economies
face AI exposure, and the OECD classifies 27\% of OECD jobs as in
occupations at high automation risk, ranging from 33.6\% (Slovak Republic)
to 18.3\% (Sweden) \citep{oecd2023employment}.

\subsection{The Productivity--Wage Divergence}
\label{sec:s_wages}

The ILO \emph{Global Wage Report 2024--25} documents that real wages in
G20 advanced economies fell 2.8\% in 2022 --- the sharpest contraction
since the series began --- before recovering to $-0.5\%$ in 2023 and
$+0.9\%$ in 2024 \citep{ilo2024wages}. Over 1999--2024 as a whole, labour
productivity in high-income countries rose 29\% while real wages rose only
15\% --- a fourteen-percentage-point cumulative divergence
\citep{ilo2024wages}.

Country-level data sharpen the picture. The U.S. labour productivity index
(BLS, business sector, 2000\,=\,100) reached 144.7 by 2024 while the real
hourly compensation index reached only 124.5 --- a twenty-percentage-point
gap \citep{bls2025productivity}. Japan presents the extreme case:
productivity rose 18\% (2000--2024) while real compensation per worker
\emph{fell} 4\% \citep{oecd2025productivity}. This wedge is visible in
Figure~\ref{fig:country_subplots} as the gap between the dashed GDP line
and the solid GAGI line in each country panel.

The labour-income-share literature tells the same story from a different
angle. \citet{karabarbounis2014global} documented a five-percentage-point
fall in the global corporate labour share between 1975 and 2012;
\citet{karabarbounis2024perspectives} confirms the decline has continued
post-2014, with the median country experiencing a further two-to-three
percentage-point decline. The U.S. business-sector labour share fell from
63.3\% (2000) to 59.1\% (2022), a 4.2-percentage-point decline
\citep{bls2025productivity}, and \citet{acemoglu2024simple} projects a
further $\sim$1 percentage-point fall over the next decade under the
current AI-deployment trajectory. On the purchasing-power side, U.S.
CPI-U rose 24.6\% between January 2019 and March 2026 while real average
hourly earnings for production and non-supervisory workers rose only 2.4\%
over the same period \citep{bls2026cpi}; in the EU-27, cumulative HICP
inflation reached 21.3\% (2019--February 2026) against essentially flat
real net wages \citep{eurostat2026hicp}; U.K. real household disposable
income per head fell 1.3\% between 2019 and 2024 \citep{ons2024hdi}; and
U.S. median household income grew 1.7\% per year in real terms
(2010--2024), compared with 2.3\% per year for the top quintile and 2.4\%
per year for the top 5\% \citep{census2024income}.

\subsection{Inequality Trends and the Policy-Regime Effect}
\label{sec:s_gini}

\begin{figure}[h]
  \centering
  \includegraphics[width=\linewidth]{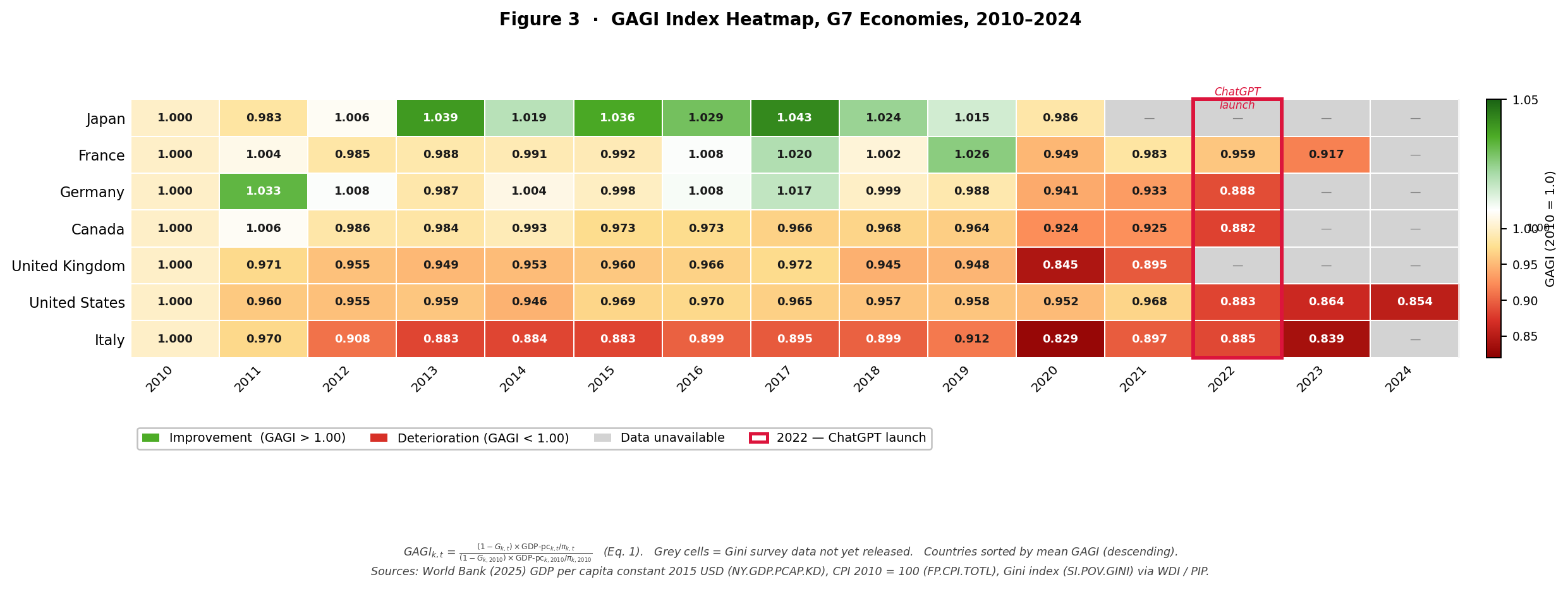}
  \caption{%
    \textbf{GAGI Index Heatmap, G7 Economies, 2010--2024.}
    Each cell shows GAGI\,=\,$(1{-}G)\times\mathrm{GDP}_{pc}/\pi$
    normalised to 2010\,=\,1.0 (Eq.~\ref{eq:gagi}); green\,=\,above
    baseline, red\,=\,below. The crimson outline marks 2022 (the
    ChatGPT launch year).
    \textbf{Observation.} Italy shows persistent red throughout the
    decade (structural stagnation); the United States deteriorated
    sharply in 2020, recovered partially, and resumed decline post-2022;
    no G7 economy sustains a strong green reading after 2022.
    \textbf{Interpretation.} The 2022 inflection visible in several
    panels is temporally coincident with the acceleration of
    AI-attributed displacement documented in \S\ref{sec:s_layoffs} ---
    an association the heatmap makes visible, though these aggregate
    series cannot, on their own, establish that automation \emph{caused}
    the inflection rather than merely co-occurring with it. Denmark and
    Sweden, two Nordic comparator economies outside the core G7 sample
    shown here (not plotted; see \citet{huf2026} for the full Nordic
    panel, computed identically), sustain green throughout the same
    period --- a difference that tracks active labour-market policy, not
    a difference in automation exposure.
    \textbf{Implication.} The heatmap converts an abstract claim about
    automation's distributional risk into a country-by-country welfare
    ledger: a regulator equipped with this metric would have been able to
    flag the post-2022 U.S. deterioration as a trigger condition while it
    was still emerging, rather than after the fact.
    Sources: \citet{worldbank2025pip,worldbank2025wdi}.%
  }
  \label{fig:gagi_heatmap}
\end{figure}

Figure~\ref{fig:gagi_heatmap} presents GAGI as a year-by-country heatmap
(green $>1.0$ = improvement; red $<1.0$ = deterioration), giving a compact
cross-sectional view of the data underlying Figure~\ref{fig:country_subplots}.

\begin{figure}[h]
  \centering
  \includegraphics[width=\linewidth]{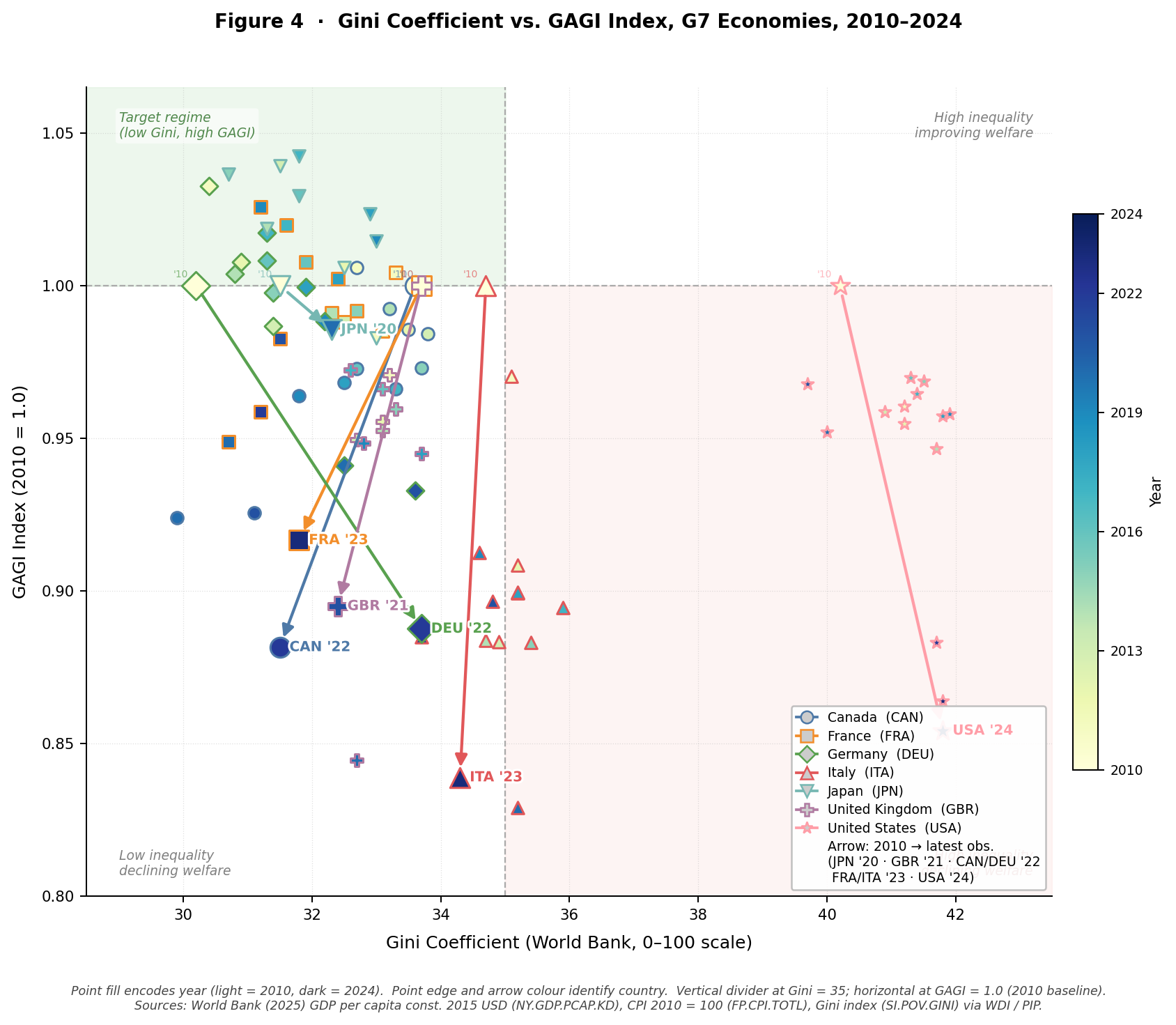}
  \caption{%
    \textbf{Gini Coefficient vs.\ GAGI Index, G7 Economies, 2010--2024.}
    Arrows connect 2010 (light) to 2024 (dark) country positions.
    Movement right\,=\,rising inequality; movement down\,=\,declining
    welfare-adjusted prosperity. The upper-left quadrant (low Gini, high
    GAGI) is the region in which automation and broadly shared prosperity
    coexist.
    \textbf{Observation.} The United States traces a predominantly
    rightward path (rising inequality with partial GAGI recovery); Italy
    traces a downward path (stagnant GAGI despite a relatively stable
    Gini coefficient); no G7 economy moves into the upper-left quadrant
    over the observation window.
    \textbf{Interpretation.} The two failure modes are qualitatively
    distinct. The U.S. failure is \emph{distributional} --- productivity
    gains are not reaching workers --- while Italy's is
    \emph{structural} --- productivity growth itself is largely absent.
    Both produce GAGI deterioration, but they call for different policy
    responses: redistribution and labour-market absorption for the United
    States, productivity-raising investment for Italy.
    \textbf{Implication.} A scalar GDP-growth figure cannot distinguish
    these two failure modes, yet the appropriate remedy is different in
    each case; the two-dimensional Gini--GAGI plane provides exactly the
    diagnostic resolution that a one-dimensional growth statistic lacks.
    Sources: \citet{worldbank2025pip,worldbank2025wdi}.%
  }
  \label{fig:gini_scatter}
\end{figure}

Figure~\ref{fig:gini_scatter} plots the Gini coefficient (horizontal axis)
against the GAGI index (vertical axis) for each G7 country, with arrows
connecting each country's 2010 position to its 2024 position.

Three contrasting regimes emerge once the G7 evidence above is read
alongside a comparator pair drawn from \emph{outside} the G7 sample ---
Denmark and Sweden --- computed identically (Eq.~\ref{eq:gagi}, same
sources) but reported separately here because neither is a G7 member; a
fuller Nordic panel appears in the companion paper \citep{huf2026}. We
include them because they are, to our knowledge, the clearest available
real-world instance of the third regime the G7 data alone cannot
illustrate: sustained automation \emph{with} absorption.

\paragraph{The Nordic comparator: automation with absorption (outside the
G7 sample).} Denmark
(Gini 26.9--28.2) and Sweden (24.1--28.4) maintain the lowest income
inequality in the OECD despite robot densities of 274 and 343 per 10,000
manufacturing employees --- the seventh- and fifth-highest in the world
\citep{ifr2024}. The mechanism is not lower automation but higher
absorption capacity: active labour-market-policy (ALMP) spending of
2.05\% of GDP in Denmark and 1.27\% in Sweden (versus an OECD average of
0.51\% and a U.S. figure of 0.10\%); unemployment-benefit replacement
rates of 64\% and 61\% (versus 38\% in the U.S.); and
collective-bargaining coverage of 82\% and 88\%
\citep{oecd2024employment,oecd2023cbcoverage,madsen2006flexicurity}. GAGI
remains stable in these countries \emph{because redistribution mechanisms
are pre-positioned}, not because automation is constrained.

\paragraph{The United States: automation without absorption.} The U.S.
presents the starkest contrast: a rising Gini coefficient
(40.0$\to$41.8 on the World Bank basis), ALMP spending of 0.10\% of GDP,
collective-bargaining coverage of 9\%, and a federal minimum wage that
fell 19\% in real terms between 2010 and 2024
\citep{dol2024mw,bls2026cpi}. AI-cited layoffs grew thirteen-fold in three
years, and workers aged 22--25 in high-exposure roles experienced a 13\%
employment decline \citep{atkinson2026young}. In
Figure~\ref{fig:country_subplots}, the U.S. panel shows the largest and
most persistent GAGI--GDP wedge among the G7.

\paragraph{The Global South: a preparedness deficit.} Only 26\% of jobs in
low-income countries are AI-exposed \citep{cazzaniga2024genai}, but the
IMF AI Preparedness Index \citep{imf2024aipi} assigns India a score of
0.49 (rank 72 of 174), Nigeria 0.32, and Bangladesh 0.38, against
Singapore's 0.80 and the United States' 0.77. The IMF's high-divergence
scenario projects between-country income inequality widening by one to
three Gini points over a decade as productivity gains concentrate in
high-preparedness economies \citep{cazzaniga2024genai}.

\subsection{Middle-Skill Hollowing}
\label{sec:s_polarization}

The share of U.S. employment in middle-wage occupations fell from 38.4\%
(2000) to 26.4\% (2024) according to the BLS Occupational Employment and
Wage Statistics \citep{bls2025oews}. In the EU-27, middle-skill employment
fell from 47.3\% (2010) to 41.1\% (2023) according to the Eurostat Labour
Force Survey \citep{eurostat2024lfs}, while the high-skill share rose from
38.5\% to 46.0\%. Occupations facing the largest projected absolute
displacement by 2030 include cashiers ($-17\%$), administrative
assistants ($-14\%$), material-recording clerks ($-11\%$), accounting and
payroll clerks ($-10\%$), data-entry clerks ($-9\%$), and bank tellers
($-5\%$) \citep{wef2025jobs}. Legal-support roles carry a 44\%
task-automatable exposure \citep{briggs2023ai}, and customer-operations
roles show the highest expected productivity substitution from generative
AI \citep{mgi2023genai}.

\section{Forward Projections Under Alternative Governance Regimes}
\label{sec:projections}

\begin{quote}
\emph{All figures in this section are model-based projections or scenario
estimates, not observed outcomes. They bound the policy stakes; they are
not point forecasts.}
\end{quote}

\citet{wef2025jobs} project 170 million new jobs created and 92 million
displaced by 2030 (a net gain of 78 million, or $+7\%$), with 22\% gross
churn across 1.2 billion formal-sector jobs and 39\% of workers' core
skills becoming obsolete or transformed. \citet{briggs2023ai} estimate
that 300 million full-time-equivalent positions globally are exposed to AI
automation, with 7\% facing full replacement and 63\% partial
transformation. \citet{mgi2023genai} estimate that 50\% of current work
activities could be automated between 2030 and 2060 (midpoint 2045) ---
a decade earlier than their 2017 forecast.

Against these relatively optimistic ranges, \citet{acemoglu2024simple}
provides a sobering upper bound: the aggregate total-factor-productivity
gain from AI over the next decade is at most 0.53--0.66\% cumulative
($\approx$0.06\% per year), with GDP effects bounded at 0.93--1.16\% over
ten years. If this bound holds, the productivity gains from AI are
themselves insufficient to fund the retraining and redistribution needed
to absorb displacement --- unless absorption mechanisms are
\emph{pre-positioned} rather than funded out of the (small) marginal
gains.

\begin{figure}[h]
  \centering
  \includegraphics[width=\linewidth]{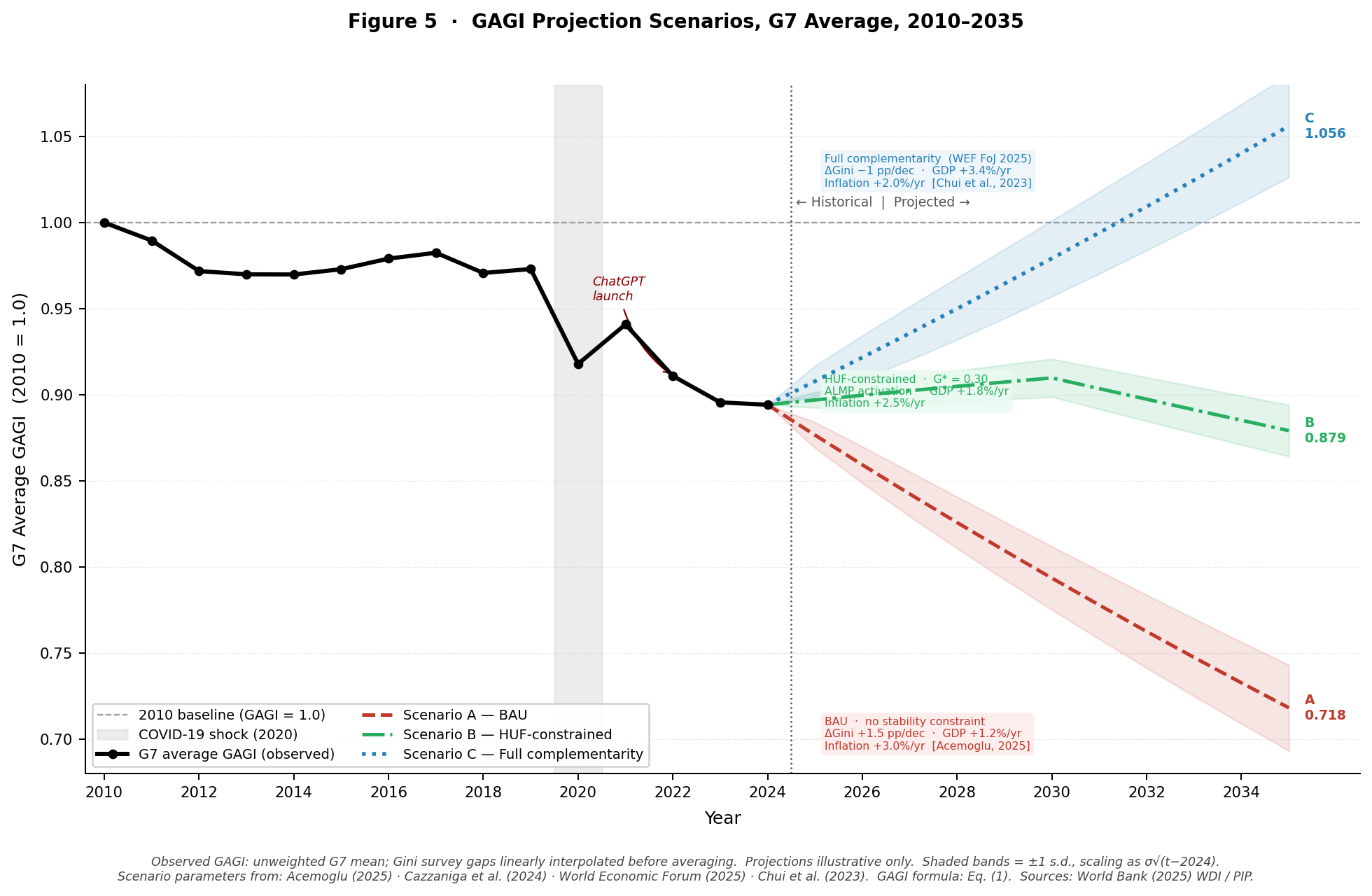}
  \caption{%
    \textbf{GAGI Projection Scenarios, G7 Average, 2010--2035.}
    \emph{All projections (2025--2035) are illustrative scenarios, not
    point forecasts; they bound the policy stakes.}
    Solid black: observed G7-average GAGI (2010--2024)
    \citep{worldbank2025pip,worldbank2025wdi}.
    Scenario A (dashed red): business-as-usual, no stability constraint;
    Gini $+1.5$ percentage points per decade; GDP growth at the Acemoglu
    ceiling \citep{acemoglu2024simple}; inflation 3\%/yr.
    Scenario B (dash-dot green): governance-constrained; Gini pinned at
    $G^{*}=0.30$ via ALMP activation; GDP growth 1.8\%/yr; inflation
    2.5\%/yr.
    Scenario C (dotted blue): full complementarity (WEF upper bound
    \citep{wef2025jobs}). Shaded bands show $\pm 1$ s.d.\ scenario
    uncertainty.
    \textbf{Observation.} Scenario A crosses a policy-defined stability
    target (here set at GAGI\,$\approx$\,0.85 for illustration; a
    governing body would determine the operative value) by the early
    2030s; Scenario B stabilises near 1.0 throughout the projection
    window.
    \textbf{Interpretation.} The five-to-ten-year separation between the
    trajectories defines a governance decision window: redistribution
    mechanisms must be pre-positioned before the target is breached,
    since \citet{korinek2019ai} show formally that Pareto-improving
    redistribution becomes infeasible once displacement exceeds a
    critical mass.
    \textbf{Implication.} Scenario B is achievable under realistic
    assumptions, but only if a governance instrument exists that can
    detect the approaching target level early enough to trigger ALMP
    activation while the option is still open.
    Sources: \citet{acemoglu2024simple,cazzaniga2024genai,%
    wef2025jobs,mgi2023genai}.%
  }
  \label{fig:gagi_projections}
\end{figure}

\paragraph{The projection model.} Figure~\ref{fig:gagi_projections}
extends Eq.~\eqref{eq:gagi} forward from the last fully observed year,
$t_0 = 2024$, by propagating its three inputs under scenario-specific
growth assumptions and recomputing GAGI from the propagated values. For
scenario $s \in \{A,B,C\}$ and year $t > t_0$,
\begin{align}
  G_{k,t}^{(s)} &=
  \begin{cases}
    G_{k,t_0} + \delta_G^{(s)} \cdot (t - t_0)
      & \text{Scenarios A, C: linear Gini drift at rate } \delta_G^{(s)} \\[2pt]
    G^{*} & \text{Scenario B: Gini pinned at } G^{*} = 0.30 \text{ via ALMP activation}
  \end{cases}
  \label{eq:proj_gini} \\[4pt]
  \mathrm{GDP\text{-}pc}_{k,t}^{(s)} &=
    \mathrm{GDP\text{-}pc}_{k,t_0} \cdot \bigl(1 + g^{(s)}\bigr)^{\,t - t_0}
  \label{eq:proj_gdp} \\[4pt]
  \pi_{k,t}^{(s)} &=
    \pi_{k,t_0} \cdot \bigl(1 + \iota^{(s)}\bigr)^{\,t - t_0}
  \label{eq:proj_cpi} \\[4pt]
  \mathrm{GAGI}_{k,t}^{(s)} &=
    \frac{\bigl(1 - G_{k,t}^{(s)}\bigr) \cdot \mathrm{GDP\text{-}pc}_{k,t}^{(s)}
          \,\big/\, \pi_{k,t}^{(s)}}
         {\bigl(1 - G_{k,2010}\bigr) \cdot \mathrm{GDP\text{-}pc}_{k,2010}
          \,\big/\, \pi_{k,2010}}
  \label{eq:proj_gagi}
\end{align}
which is simply Eq.~\eqref{eq:gagi} evaluated at the propagated inputs
rather than at observed ones; the G7-average curve plotted in
Figure~\ref{fig:gagi_projections} is the simple cross-country mean of
$\mathrm{GAGI}_{k,t}^{(s)}$ over $k \in \{\text{G7}\}$. The
scenario-specific parameter triple $(\delta_G^{(s)}, g^{(s)}, \iota^{(s)})$
is fixed exogenously from the literature ranges discussed above and
reported in full in the figure caption: Scenario A (business-as-usual)
sets $\delta_G^{(A)} = +0.15$ Gini points/year (i.e., $+1.5$ points/decade),
$g^{(A)}$ at the \citet{acemoglu2024simple} GDP-growth ceiling, and
$\iota^{(A)} = 3\%$/year; Scenario B (governance-constrained) replaces the
Gini-drift term with the pinned value $G^{*}$, sets $g^{(B)} = 1.8\%$/year,
and $\iota^{(B)} = 2.5\%$/year; Scenario C (full complementarity) adopts
the \citet{wef2025jobs} upper-bound employment-and-skills assumptions in
place of $g^{(C)}$. We emphasise that Eqs.~\eqref{eq:proj_gini}--%
\eqref{eq:proj_gagi} are an explicit \emph{accounting identity applied to
assumed inputs}, not an estimated structural model: the parameters are
scenario assumptions drawn from the cited literature ranges, not fitted
coefficients, and the resulting curves should be read as bounding
illustrations of policy stakes (as the figure's disclaimer states), not as
point forecasts of what will occur.

\citet{cazzaniga2024genai} simulate that, under a high-displacement,
low-redistribution scenario analogous to Scenario A, income inequality in
advanced economies rises by 0.5--2 Gini points within a decade, with
wealth inequality widening more robustly across all scenarios.
\citet{korinek2019ai} show formally that Pareto improvements from AI
require redistribution mechanisms to be activated \emph{before} the
displacement threshold is crossed, and \citet{acemoglu2019automation}
demonstrate that ``automation traps'' --- equilibria in which excessive
displacement reduces aggregate output below the pre-automation baseline
--- are theoretically possible under plausible parameterisations.

\section{Discussion}
\label{sec:discussion}

\paragraph{What GAGI adds to existing welfare metrics.} Inequality-adjusted
income measures are not new: the Atkinson index, the inequality-adjusted
Human Development Index, and the World Bank's shared-prosperity premium
all combine growth and distribution in various ways. GAGI's contribution
is not conceptual novelty but \emph{fitness for a specific governance
purpose}: it is computable annually, for any country with World Bank
data coverage, from three widely published series, normalised in a way
that makes year-on-year and cross-country \emph{change} comparable. This
combination --- minimal data requirements, transparent construction, and
a fixed-baseline normalisation --- is what allows GAGI to function as a
monitoring trigger rather than a retrospective research statistic. We see
this as the central practical argument for adding a GAGI-type index to
the standard macroeconomic dashboard alongside GDP, the employment
ratio, and inflation.

\paragraph{Toward a distribution-aware misery indicator.} A related
opportunity lies in how GAGI could improve on the conventional
\emph{misery index} --- Okun's sum of the unemployment and inflation
rates, and its later variants that add lending rates or subtract GDP
growth. That index is a purely macro-aggregate construction: it
implicitly treats each percentage point of unemployment or inflation as
imposing the same hardship on every household, which amounts to assuming
that income, wealth, and the cost of living are distributed uniformly
across the population. This assumption fails broadly --- not only in
the context of AI-driven displacement but in any economy in which job
losses, productivity gains, and cost-of-living pressures are
distributed unevenly. A given inflation rate erodes welfare far more in
a high-Gini, high-cost-of-living region than in a low-Gini,
low-cost-of-living one, and displacement concentrated in particular
occupation categories (\S\ref{sec:s_polarization}) imposes far more
aggregate hardship per statistical job-loss than displacement spread
uniformly. Because GAGI already combines distribution ($1-G$) and an
aggregate price adjustment ($\pi$) into a single welfare-adjusted
prosperity measure, the ratio of GAGI to a region- or country-specific
cost-of-living index ---
\[
  \mathrm{GAGI}_{k,t} \;/\; \mathrm{COL}_{k,t},
  \quad\text{or its inverse}\quad
  \mathrm{COL}_{k,t} \;/\; \mathrm{GAGI}_{k,t}
\]
read as a misery proxy --- is a natural, distribution-aware
alternative. Where the conventional misery index asks ``how
uncomfortable is the macroeconomy on average?'', this ratio asks the
more policy-relevant question ``how far does a typical,
inequality-adjusted unit of income actually stretch against the cost of
living faced where people live?''

This ratio becomes considerably more interpretable when read alongside
the \emph{employment ratio} --- defined here as the share of the
working-age population that is purposefully and productively employed,
explicitly excluding not only those classified as unemployed but also
the underemployed (involuntary part-time workers and workers in roles
substantially below their productive capacity) and those who have
withdrawn from the labour market entirely. The conventional
unemployment rate omits these latter two groups and therefore
understates the true shortfall in productive labour utilisation; the
employment ratio as defined above does not. Jointly, GAGI/COL and the
employment ratio would provide a compact, two-dimensional picture of
whether prosperity is welfare-adjusted and broadly accessible (GAGI/COL
high) and whether the labour market is genuinely absorbing the working-age
population (employment ratio high) --- a summary that is notably more
informative than the conventional misery index without requiring
additional modelling assumptions.

We flag both the GAGI/COL ratio and the accompanying employment-ratio
framing as promising empirical extensions rather than fully developed
instruments in this paper. Building GAGI/COL at sub-national resolution
would require regional cost-of-living series that are far less
consistently published than the national CPI series underlying
Eq.~\eqref{eq:gagi}; constructing a cross-country employment-ratio
series using the definition above would require harmonising
underemployment and labour-force-withdrawal estimates across statistical
agencies. Both are feasible with existing microdata and are left to
future work; the conceptual case for the combination is established here.

\paragraph{The stability-threshold argument.} The patterns documented in
\S\ref{sec:s_country}--\S\ref{sec:s_polarization} support a structural
claim: there exists a combination of Gini, labour-force participation,
and real purchasing power beyond which an economic system loses the
capacity to self-correct through market mechanisms alone --- tax revenues
fall, retraining budgets contract, political stability deteriorates, and
the social compact that enables long-term investment erodes. The GAGI
value at which this transition occurs is not empirically derived from
the data in this paper; it is a policy-settable target that a governing
body would calibrate to its own economic and social priorities. The
value GAGI\,$\approx$\,0.85 used in Figure~\ref{fig:gagi_projections}
is purely indicative --- chosen to illustrate the decision-window
argument, not to assert a universal threshold. Figure~\ref{fig:gagi_projections}
shows that a business-as-usual trajectory (Scenario A) reaches this
indicative target by the early 2030s under plausible parameters, while
a governance-constrained trajectory (Scenario B) avoids it. The absence of
a formal stability constraint in current macroeconomic monitoring and
automation-governance instruments is therefore not merely an
architectural gap: it is the absence of an early-warning system for a
risk that is already accumulating in the data presented here.

\paragraph{Limitations.} GAGI inherits the limitations of its inputs.
Gini coefficients are measured with lags and methodological
inconsistencies across countries \citep{worldbank2025pip}; GDP-per-capita
and CPI series are revised retrospectively; and the index, like any
scalar, compresses a multi-dimensional distributional reality into a
single number. GAGI is also a \emph{relative} measure --- it tracks change
from a 2010 baseline and says nothing about absolute welfare levels, so
it should be read alongside (not instead of) level statistics such as
absolute poverty rates and median income. Finally, the 2025--2026 values
used here are preliminary or partial-year estimates and should be treated
as provisional until finalised data are released.

\paragraph{Relationship to the Human Utility Factor.} GAGI was developed
as an empirical proxy for the Economic Stability component of the Human
Utility Factor (HUF), a derived, three-component welfare metric
($\mathrm{HUF} = A \times W \times E$, where $A$ is Agency, $W$ is
Wellbeing, and $E$ is Economic Stability) that reframes the governance
of automation as a constrained-optimisation problem: it identifies the
parameters --- automation depth, redistribution intensity, and
employment coverage --- that must be tuned jointly to achieve
macro-socio-economic alignment \citep{huf2026}. GAGI operationalises
the $E$ component of that framework, providing a publicly computable,
annually trackable signal of whether the economic-stability dimension of
wellbeing is moving in the right direction. The empirical patterns
documented here --- the persistent GAGI--GDP wedge, the acceleration of
displacement after 2022, and the separation of national trajectories by
policy regime --- provide the empirical motivation for that framework:
they show that the distributional misalignments HUF is designed to
detect and correct are not hypothetical, but already measurable and
accelerating in the historical record.

\section{Conclusion}
\label{sec:conclusion}

GDP-based monitoring cannot detect the welfare cost of automation-driven
distributional harm, because by construction it averages over the very
distributional dimension in which that harm appears. We have introduced
GAGI, a simple and reproducible alternative that rescales GDP per capita
by inequality and inflation, and applied it to the G7 economies over
2010--2026. The result is a welfare ledger that is invisible in GDP terms:
a persistent and widening gap between headline growth and
welfare-adjusted prosperity, an acceleration of that gap that is
temporally coincident with --- though not, on this evidence alone, shown
to be caused by --- the period of fastest AI deployment, and a clean
separation of regimes: automation-without-absorption (United States) and
stagnation-without-automation (Italy, Japan) within the core G7 sample,
contrasted with automation-with-absorption in Nordic comparator economies
drawn from outside the G7 (Denmark, Sweden). None of these patterns require speculative
modelling to observe --- they are visible in publicly available statistics
once the right lens is applied. We offer GAGI as that lens, and as a candidate building block for any
macroeconomic monitoring or automation-governance instrument that
intends to track \emph{welfare}, rather than merely \emph{output}, as
automation --- of whatever kind and at whatever pace --- proceeds.

\section*{Acknowledgements}

The author thanks Dr.\ Gilberto Ochoa for his arXiv endorsement, which
made it possible to submit this work to the public preprint record.

\bibliographystyle{plainnat}
\bibliography{references(1)}

\end{document}